\documentclass[conference,10pt]{IEEEtran}
\usepackage{url}
\usepackage[utf8]{inputenc}
\usepackage{color}
\usepackage{amsmath}
\usepackage{amssymb}

\usepackage[acronyms,nonumberlist,nopostdot,nomain,nogroupskip]{glossaries}
\usepackage{tablefootnote}
\usepackage{booktabs}

\usepackage{tikz}
\usepackage{pgfplots}
\pgfplotsset{compat=newest} 
\pgfplotsset{plot coordinates/math parser=false} 
\newlength\fheight
\newlength\fwidth
\usetikzlibrary{plotmarks,patterns,decorations.pathreplacing,backgrounds,calc,arrows,arrows.meta,spy,matrix}
\usepgfplotslibrary{patchplots,groupplots}
\usepackage{tikzscale}

\usepackage{multirow}
\usepackage{tkz-kiviat}

\usepackage[font=scriptsize]{subcaption}
\usepackage[font=footnotesize]{caption}

\usepackage{mathtools}

\newacronym{3gpp}{3GPP}{3rd Generation Partnership Project}
\newacronym{adc}{ADC}{Analog to Digital Converter}
\newacronym{5g}{5G}{5th generation}
\newacronym{aimd}{AIMD}{Additive Increase Multiplicative Decrease}
\newacronym{am}{AM}{Acknowledged Mode}
\newacronym{amc}{AMC}{Adaptive Modulation and Coding}
\newacronym{aqm}{AQM}{Active Queue Management}
\newacronym{awgn}{AGWN}{Additive White Gaussian Noise}
\newacronym{balia}{BALIA}{Balanced Link Adaptation}
\newacronym{bdp}{BDP}{Bandwidth-Delay Product}
\newacronym{bf}{BF}{Beamforming}
\newacronym{cc}{CC}{Congestion Control}
\newacronym{cdf}{CDF}{Cumulative Distribution Function}
\newacronym{cn}{CN}{Core Network}
\newacronym{cqi}{CQI}{Channel Quality Information}
\newacronym{cp}{CP}{Control Plane}
\newacronym{csirs}{CSI-RS}{Channel State Information - Reference Signal}
\newacronym{dc}{DC}{Dual Connectivity}
\newacronym{dce}{DCE}{Direct Code Execution}
\newacronym{dci}{DCI}{Downlink Control Information}
\newacronym{dl}{DL}{Downlink}
\newacronym{dmr}{DMR}{Deadline Miss Ratio}
\newacronym{dmrs}{DMRS}{DeModulation Reference Signal}
\newacronym{e2e}{E2E}{End-to-End}
\newacronym{ecn}{ECN}{Explicit Congestion Notification}
\newacronym{edf}{EDF}{Earliest Deadline First}
\newacronym{enb}{eNB}{evolved Node Base}
\newacronym{epc}{EPC}{Evolved Packet Core}
\newacronym{es}{ES}{Edge Server}
\newacronym{fdma}{FDMA}{Frequency Division Multiple Access}
\newacronym{fdd}{FDD}{Frequency Division Duplexing}
\newacronym[firstplural=Radio Access Technologies (RATs)]{rat}{RAT}{Radio Access Technology}
\newacronym{fs}{FS}{Fast Switching}
\newacronym{ftp}{FTP}{File Transfer Protocol}
\newacronym{gnb}{gNB}{Next Generation Node Base Station}
\newacronym{harq}{HARQ}{Hybrid Automatic Repeat reQuest}
\newacronym{hetnet}{HetNet}{Heterogeneous Network}
\newacronym{hh}{HH}{Hard Handover}
\newacronym{hol}{HOL}{Head-of-Line}
\newacronym{ia}{IA}{Initial Access}
\newacronym{imt}{IMT}{International Mobile Telecommunication}
\newacronym{iot}{IoT}{Internet of Things}
\newacronym{los}{LOS}{Line of Sight}
\newacronym{lte}{LTE}{Long Term Evolution}
\newacronym{m2m}{M2M}{Machine to Machine}
\newacronym{mac}{MAC}{Medium Access Control}
\newacronym{mc}{MC}{Multi-Connectivity}
\newacronym{mcs}{MCS}{Modulation and Coding Scheme}
\newacronym{mec}{MEC}{Mobile Edge Cloud}
\newacronym{mi}{MI}{Mutual Information}
\newacronym{mimo}{MIMO}{Multiple Input, Multiple Output}
\newacronym{mmwave}{mmWave}{millimeter wave}
\newacronym{mptcp}{MPTCP}{Multipath TCP}
\newacronym{mr}{MR}{Maximum Rate}
\newacronym{mss}{MSS}{Maximum Segment Size}
\newacronym{mtd}{MTD}{Machine-Type Device}
\newacronym{mtu}{MTU}{Maximum Transmission Unit}
\newacronym{nfv}{NFV}{Network Function Virtualization}
\newacronym{nlos}{NLOS}{Non Line of Sight}
\newacronym{nr}{NR}{NR}
\newacronym{ofdm}{OFDM}{Orthogonal Frequency Division Multiplexing}
\newacronym{pdcch}{PDCCH}{Physical Downlonk Control Channel}
\newacronym{pdcp}{PDCP}{Packet Data Convergence Protocol}
\newacronym{pdsch}{PDSCH}{Physical Downlink Shared Channel}
\newacronym{pdu}{PDU}{Packet Data Unit}
\newacronym{pf}{PF}{Proportional Fair}
\newacronym{pgw}{PGW}{Packet Gateway}
\newacronym{phy}{PHY}{Physical}
\newacronym{pbch}{PBCH}{Physical Broadcast Channel}
\newacronym[plural=\gls{mme}s,firstplural=Mobility Management Entities (MMEs)]{mme}{MME}{Mobility Management Entity}
\newacronym{prb}{PRB}{Physical Resource Block}
\newacronym{pss}{PSS}{Primary Synchronization Signal}
\newacronym{pucch}{PUCCH}{Physical Uplink Control Channel}
\newacronym{pusch}{PUSCH}{Physical Uplink Shared Channel}
\newacronym{rach}{RACH}{Random Access Channel}
\newacronym{ran}{RAN}{Radio Access Network}
\newacronym{red}{RED}{Random Early Detection}
\newacronym{rf}{RF}{Radio Frequency}
\newacronym{rlc}{RLC}{Radio Link Control}
\newacronym{rlf}{RLF}{Radio Link Failure}
\newacronym{rrc}{RRC}{Radio Resource Control}
\newacronym{rrm}{RRM}{Radio Resource Management}
\newacronym{rr}{RR}{Round Robin}
\newacronym{rs}{RS}{Remote Server}
\newacronym{rsrp}{RSRP}{Reference Signal Received Power}
\newacronym{rss}{RSS}{Received Signal Strength}
\newacronym{rtt}{RTT}{Round Trip Time}
\newacronym{rw}{RW}{Receive Window}
\newacronym{rx}{RX}{Receiver}
\newacronym{sa}{SA}{Standalone}
\newacronym{sack}{SACK}{Selective Acknowledgment}
\newacronym{sap}{SAP}{Service Access Point}
\newacronym{sch}{SCH}{Secondary Cell Handover}
\newacronym{scoot}{SCOOT}{Split Cycle Offset Optimization Technique}
\newacronym{sdma}{SDMA}{Spatial Division Multiple Access}
\newacronym{sinr}{SINR}{Signal to Interference plus Noise Ratio}
\newacronym{sm}{SM}{Saturation Mode}
\newacronym{snr}{SNR}{Signal to Noise Ratio}
\newacronym{son}{SON}{Self-Organizing Network}
\newacronym{ss}{SS}{Synchronization Signal}
\newacronym{srs}{SRS}{Sounding Reference Signal}
\newacronym{sss}{SSS}{Secondary Synchronization Signal}
\newacronym{tb}{TB}{Transport Block}
\newacronym{tcp}{TCP}{Transmission Control Protocol}
\newacronym{tdd}{TDD}{Time Division Duplexing}
\newacronym{tdma}{TDMA}{Time Division Multiple Access}
\newacronym{tfl}{TfL}{Transport for London}
\newacronym{tm}{TM}{Transparent Mode}
\newacronym{trp}{TRP}{Transmitter Receiver Pair}
\newacronym{tti}{TTI}{Transmission Time Interval}
\newacronym{ttt}{TTT}{Time-to-Trigger}
\newacronym{tx}{TX}{Transmitter}
\newacronym{ue}{UE}{User Equipment}
\newacronym{ul}{UL}{Uplink}
\newacronym{uml}{UML}{Unified Modeling Language}
\newacronym{um}{UM}{Unacknowledged Mode}
\newacronym{utc}{UTC}{Urban Traffic Control}
\newacronym{vm}{VM}{Virtual Machine}
\newacronym{rsrq}{RSRQ}{Reference Signal Received Quality}
\newacronym{rssi}{RSSI}{Received Signal Strength Indicator}
\newacronym{crs}{CRS}{Cell Reference Signal}
\newacronym{nsa}{NSA}{Non-Standalone}
\newacronym{mrdc}{MR-DC}{Multi \gls{rat} \gls{dc}}
\newacronym{endc}{EN-DC}{E-UTRAN-\gls{nr} \gls{dc}}
\newacronym{5gc}{5GC}{5G Core}
\tikzstyle{startstop} = [rectangle, rounded corners, minimum width=2cm, minimum height=0.5cm,text centered, draw=black]
\tikzstyle{io} = [trapezium, trapezium left angle=70, trapezium right angle=110, minimum width=3cm, minimum height=1cm, text centered, draw=black]
\tikzstyle{process} = [rectangle, minimum width=2cm, minimum height=0.5cm, text centered, draw=black, alignb=center]
\tikzstyle{decision} = [ellipse, minimum width=2cm, minimum height=1cm, text centered, draw=black]
\tikzstyle{arrow} = [thick,<->,>=stealth]
\tikzstyle{line} = [thick,>=stealth]
\tikzstyle{darrow} = [thick,<->,>=stealth,dashed]
\tikzstyle{sarrow} = [thick,->,>=stealth]
\tikzstyle{larrow} = [line width=0.1mm,dashdotted,->,>=stealth]

\makeatletter
\def\grd@save@target#1{%
  \def\grd@target{#1}}
\def\grd@save@start#1{%
  \def\grd@start{#1}}
\tikzset{
  grid with coordinates/.style={
    to path={%
      \pgfextra{%
        \edef\grd@@target{(\tikztotarget)}%
        \tikz@scan@one@point\grd@save@target\grd@@target\relax
        \edef\grd@@start{(\tikztostart)}%
        \tikz@scan@one@point\grd@save@start\grd@@start\relax
        \draw[minor help lines] (\tikztostart) grid (\tikztotarget);
        \draw[major help lines] (\tikztostart) grid (\tikztotarget);
        \grd@start
        \pgfmathsetmacro{\grd@xa}{\the\pgf@x/1cm}
        \pgfmathsetmacro{\grd@ya}{\the\pgf@y/1cm}
        \grd@target
        \pgfmathsetmacro{\grd@xb}{\the\pgf@x/1cm}
        \pgfmathsetmacro{\grd@yb}{\the\pgf@y/1cm}
        \pgfmathsetmacro{\grd@xc}{\grd@xa + \pgfkeysvalueof{/tikz/grid with coordinates/major step x}}
        \pgfmathsetmacro{\grd@yc}{\grd@ya + \pgfkeysvalueof{/tikz/grid with coordinates/major step y}}
        \foreach \x in {\grd@xa,\grd@xc,...,\grd@xb}
        \node[anchor=north] at (\x,\grd@ya) {\pgfmathprintnumber{\x}};
        \foreach \y in {\grd@ya,\grd@yc,...,\grd@yb}
        \node[anchor=east] at (\grd@xa,\y) {\pgfmathprintnumber{\y}};
      }
    }
  },
  minor help lines/.style={
    help lines,
    gray,
    line cap =round,
    xstep=\pgfkeysvalueof{/tikz/grid with coordinates/minor step x},
    ystep=\pgfkeysvalueof{/tikz/grid with coordinates/minor step y}
  },
  major help lines/.style={
    help lines,
    line cap =round,
    line width=\pgfkeysvalueof{/tikz/grid with coordinates/major line width},
    xstep=\pgfkeysvalueof{/tikz/grid with coordinates/major step x},
    ystep=\pgfkeysvalueof{/tikz/grid with coordinates/major step y}
  },
  grid with coordinates/.cd,
  minor step x/.initial=.5,
  minor step y/.initial=.2,
  major step x/.initial=1,
  major step y/.initial=1,
  major line width/.initial=1pt,
}
\makeatother

\makeglossaries

\IEEEoverridecommandlockouts
\IEEEpubid{%
\makebox[\columnwidth]{ISBN 978-3-903176-05-8~\copyright~2018 IFIP \hfill}%
\hspace{\columnsep}%
\makebox[\columnwidth]{ }%
}

\begin{document}

	
\title{Initial Access Frameworks for\\3GPP NR at mmWave Frequencies\vspace{-0.3cm}}

\author{\IEEEauthorblockN{Marco Giordani$^{\circ }$, Michele Polese$^{\circ }$, Arnab Roy$^{\dagger }$, Douglas Castor$^{\dagger }$, Michele Zorzi$^{\circ }$}
\IEEEauthorblockA{
\small email:\texttt{\{giordani,polesemi,zorzi\}@dei.unipd.it, }
\texttt{\small \{arnab.roy,douglas.castor\}@interdigital.com}\\
$^{\circ }$\small Consorzio Futuro in Ricerca (CFR)   University of Padova, Italy \qquad
$^{\dagger }$\small InterDigital Communications, Inc., USA\\}}

\makeatletter
\patchcmd{\@maketitle}
  {\addvspace{0.5\baselineskip}\egroup}
  {\addvspace{-1.1\baselineskip}\egroup}
  {}
  {}
\makeatother
\flushbottom
\setlength{\parskip}{0ex plus0.1ex}

\maketitle

\begin{abstract}
The  use of millimeter wave (mmWave) frequencies for communication will be one of the innovations of the next generation of cellular mobile networks (5G). It will provide unprecedented data rates, but is highly susceptible to rapid channel variations and suffers from severe isotropic pathloss. Highly directional antennas at the transmitter and the receiver will be used to compensate for these shortcomings and achieve sufficient link budget in wide area networks. 
However, directionality demands precise alignment of the transmitter and the receiver beams, an operation which has important implications for control plane procedures, such as initial access, and may increase the delay of the data transmission.
This paper provides a comparison of measurement frameworks for initial access in mmWave cellular networks in terms of detection accuracy, reactiveness and overhead, using parameters recently standardized by the 3GPP and a channel model based on real-world measurements.
We show that the best strategy depends on the specific environment in which the nodes are deployed, and  provide guidelines to characterize the optimal choice as a function of the system~parameters.
\end{abstract}


\vspace{-.2cm}
\section{Introduction}
\vspace{-.1cm}
The \gls{5g} of mobile cellular networks is currently being standardized by the 3GPP as NR~\cite{38300}, and is designed to enable
a fully mobile and connected society, in order to address the tremendous growth in connectivity and density/volume of traffic that will be required in the near future \cite{boccardi2014five}.
 In particular, 5G cellular networks are designed to provide very high throughput (1 Gbps or more), ultra-low latency (even less than 1 ms in some cases), ultra-high reliability, low energy consumption, and ultra-high connectivity resilience~\cite{boccardi2014five}.

The mmWave spectrum -- roughly comprised between  10 and 300 GHz
 -- has been considered as an enabler of the 5G performance requirements in micro and
picocellular networks~\cite{rangan2017potentials}.
These frequencies offer much more bandwidth than current cellular systems, which are allocated in the congested bands below 6 GHz, and initial capacity estimates have suggested that mmWave networks offer orders of magnitude higher bit-rates than 4G systems~\cite{akoum2012coverage}.
However, the increased carrier frequency makes the propagation conditions more demanding than at the lower frequencies traditionally used for wireless services, making the communication less reliable~\cite{pi2011introduction}. 
MmWave signals suffer from a higher pathloss and severe channel intermittency, and are blocked by many common materials~\cite{lu2012modeling}. As a result, the quality of the wireless link between the \gls{ue} and the network can be highly variable.

To overcome these limitations, next-generation cellular systems will be equipped with high-dimensional phased arrays both at the \gls{ue} and at the \gls{gnb}, in order to establish highly directional transmission links and to benefit from the resulting beamforming gain. Directional links, however, require fine alignment of the transmitter and the receiver beams, a procedure which might greatly increase the time it takes to access the network. 
In this regard, defining efficient \gls{ia} procedures, which allow a mobile \gls{ue} to establish a physical link connection with a \gls{gnb} (a necessary step to access the network), is particularly challenging at mmWave frequencies~\cite{giordani2016initial}.
In current \gls{lte} systems, \gls{ia} is performed on omnidirectional channels, whereas beamforming or other directional transmissions can only be performed after a physical link is established~\cite{36300}. 
On the other hand, in the mmWave bands, it may  be essential to exploit the antenna gains already during the \gls{ia} phase, otherwise there would be a mismatch between the range at which a cell can be detected (control-plane range), and the much longer range at which a user could directionally send and receive data using beamforming (user-plane range).

\subsection{Related Work}

Papers on \gls{ia} (e.g., \cite{ giordani2016initial, liu2016user}) in \gls{5g} mmWave cellular systems are very recent.\footnote{For a complete overview of the most relevant works on  beam management we refer to \cite{giordani2018tutorial}.}
Most literature refers to lower frequencies in ad hoc wireless network scenarios or, more recently, to the 60 GHz IEEE 802.11ad WLAN and WPAN scenarios (e.g., \cite{nitsche201460ghz}).
However, most of the proposed solutions present many limitations (e.g., they
are appropriate for short-range, static and indoor scenarios) which prevent them from matching the requirements of  cellular systems.

Recently, new solutions specifically designed for mobile wireless networks have been proposed.
In \cite{jeong2015random, barati2015directional}, the authors proposed an exhaustive method that performs \gls{ia} over mmWave frequencies by periodically transmitting directional synchronization signals to scan the angular space. With this approach, a large number of antennas at the transceiver makes it possible to reach more users with respect to the case with a single omnidirectional antenna, at the cost of a longer delay for the \gls{ia}. 
Additionally, more sophisticated discovery techniques (e.g., \cite{desai2014initial}) decrease the exhaustive search delay through a multi-phase hierarchical procedure, in which access signals are first sent over a few directions, with wide beams, and then iteratively refined until the beamforming gain is sufficiently high.
In \cite{choi2015beam} a low-complexity beam selection method is derived by exploiting the sparsity of mmWave channels, and thanks to compressive sensing it is possible to avoid explicit channel estimation for beam management.

The performance of the  association techniques also depends on the beamforming architecture implemented in the transceivers.
Preliminary works aiming at finding the optimal beamforming strategy refer to WLAN scenarios. For
example, the algorithm proposed in \cite{chandra2014adaptive} takes into account the spatial distribution of nodes, to allocate the beamwidth of each antenna pattern in an adaptive fashion and satisfy the required link budget criterion.
Since the proposed algorithm minimizes the collisions, it also minimizes the average time required to transmit a data packet from the source to the destination through a specific direction.
In \gls{5g} scenarios, papers \cite{jeong2015random, barati2015directional,desai2014initial} give some insights on tradeoffs among different beamforming architectures in terms of user association's quality.

\vspace{-.2cm}
\subsection{Contributions of This Paper}
\vspace{-.1cm}
In this paper we provide the first global comprehensive evaluation of mmWave measurement frameworks for \gls{ia}, using 3GPP NR scenario configurations (e.g.,  the NR frame structure and other relevant physical-layer parameters), and assess how to optimally design fast, accurate and robust control-plane management schemes through measurement reports.
We focus on \gls{dl} and \gls{ul} frameworks, and on \gls{sa} and \gls{mc} architectures. 
We simulate their performance in terms of (i) \emph{detection accuracy}, i.e., how representative the measurement is; (ii) \emph{reactiveness}, i.e., how quickly a mobile user gets access to the network; and (iii) \emph{overhead}, i.e., how many resources are needed for the measurement operations. Finally, we illustrate some of the complex tradeoffs to be considered when designing \gls{ia} solutions for 3GPP NR.
The results prove that the optimal design for implementing efficient and fast \gls{ia}
must
account for several specific features such as the \glspl{gnb} density, the antenna geometry, the beamforming configuration and the level of integration and harmonization of different~technologies.

The rest of the paper is organized as follows. 
In Sec. \ref{sec:meas_frameworks} we provide an introduction on 3GPP NR procedures for \gls{ia} at mmWave frequencies, and present the \gls{ia} frameworks we will evaluate. 
In Sec. \ref{sec:system_model} we describe the system model, the metrics that will be considered and the 3GPP parameters that will be configured.
In Sec. \ref{sec:results} we present our main findings and results, while Sec. \ref{sec:concl} concludes the paper.

\section{Initial Access Frameworks}
\label{sec:meas_frameworks}

3GPP NR will support a wide range of frequencies and use cases, and is designed to support beamforming operations for both data and control planes~\cite{38300} in the \gls{phy} and \gls{mac} layers.
In particular, for \gls{ia}, the concept of \gls{ss} block and burst recently emerged for the periodic transmission of synchronization signals from the \glspl{gnb}~\cite{38211}. These signals can be used at the receiver side to estimate the channel and select the best \gls{gnb} to attach to. 
An \gls{ss} block is a group of 4 \gls{ofdm} symbols in time and 240 subcarriers in frequency~\cite{38211}, and carries the \gls{pss}, the \gls{sss} and the \gls{pbch}. The \gls{dmrs} associated with the \gls{pbch} can be used to estimate the \gls{rsrp} of each \gls{ss} block. For each slot of 14 symbols there can be up to two \gls{ss} blocks~\cite{38211}. The \gls{ss} blocks are grouped in an \gls{ss} burst, which lasts up to 5 ms, and can be repeated after $T_{\rm SS} \in \{5, 10, 20, 40, 80, 160\}$ ms~\cite{38331}.
The maximum number $L$ of \gls{ss} blocks in a burst is frequency-dependent~\cite{38211}, and above 6 GHz there could be up to 64 blocks per burst. 
When considering frequencies for which beam operations are required, each \gls{ss} block is mapped to a certain angular direction. 

Moreover, 3GPP NR specifications include a set of basic procedures for beam management~\cite{38300}. We analyze two different deployment architectures. With the \textit{standalone} option, the \gls{ue} connects only to an NR \gls{gnb} at mmWave frequencies. With \textit{multi-connectivity}, instead, each \gls{ue} maintains multiple possible signal paths to different cells at different frequencies (e.g., NR at mmWave and LTE at conventional frequencies), thus providing both high capacity and robust connections~\cite{polese2017jsac}. We also distinguish between a \textit{downlink} and an \textit{uplink} framework. 
In the downlink case, the \glspl{gnb} transmit synchronization signals (i.e., \gls{ss} blocks) which are collected by the surrounding \glspl{ue}, while in the uplink case the measurements are based on \glspl{srs} transmitted by the mobile terminals.
Notice that the 3GPP considers only the downlink framework for \gls{ia}. Nonetheless, it is worth comparing the downlink and uplink solutions, given that the rising heterogeneity in cellular networks is dramatically changing the traditional notion of a communication cell~\cite{boccardi2014five}, increasing the importance of the uplink traffic and advocating the design of \gls{ul}-driven solutions for both the data and the control planes.

The first procedure for \gls{ia} is \textit{beam sweeping}, 
 i.e., covering a spatial area with a set of beams transmitted and received according to pre-specified intervals and directions. 
The second procedure, denominated \textit{beam measurement}, requires the \glspl{ue} in a downlink framework (or the \glspl{gnb} in an uplink one) to evaluate the quality of the received signal. 
Different metrics could be used~\cite{38215}. In this paper, we consider the \gls{snr}, which is the linear average of the received power on different resources with synchronization signals divided by the noise power.
The third procedure is \textit{beam determination}, i.e., the selection of the best beams  at the \gls{gnb} and at the \gls{ue}, according to the measurements obtained with the beam measurement procedure. This procedure differs in the \gls{dl} and \gls{ul} frameworks. In downlink, the \gls{ue} performs autonomously a decision on the best direction in which \gls{ia} should be performed. In uplink, instead, the \glspl{gnb} forward the  channel measurements to a central controller, which then decides which is the best direction.\footnote{We recall that the optimal beam pair for each link can be determined only after a complete scan, since the  \glspl{gnb} have to  detect all \glspl{ue} within their whole angular range.}
The fourth and final procedure is \textit{beam reporting}, i.e., information on the quality of the received beamformed signals and on the decisions in the beam determination phase is exchanged. In a standalone downlink (SA-DL) framework, as proposed by the 3GPP, the mobile terminal has to wait for the \gls{gnb} to schedule the \gls{rach} opportunity towards the best direction that the \gls{ue} has just determined, for performing random access and implicitly informing the selected serving infrastructure about the optimal direction through which it has to steer its beam. As reported in~\cite{ericsson2017rach}, it has been agreed that for each \gls{ss} block the \gls{gnb} will specify one or more \gls{rach} opportunities with a certain time and frequency offset and direction, so that the \gls{ue} knows when to transmit the \gls{rach} preamble. When a multi-connectivity downlink (MC-DL) scheme is considered, instead, the \gls{ue} can use the \gls{lte} connection to report the optimal set of directions to the \glspl{gnb}, so that it does not need to wait for an additional beam sweep from the \gls{gnb} to perform the beam reporting or the \gls{ia} procedure. 
Similarly, in a multi-connectivity uplink (MC-UL) framework, the network reports to the \gls{ue} the optimal direction and the resources for random access. Notice that we do not consider the \gls{sa}-\gls{ul} configuration for \gls{ia}, since we believe that uplink-based architectures will likely necessitate the support of an \gls{lte} overlay for the management of the control plane and the implementation of efficient measurement operations.

\section{System Model}
\label{sec:system_model}

\subsection{Performance Metrics}
The performance of the different architectures and beam management procedures for \gls{ia} will be assessed using three different metrics. The \textit{detection accuracy} is measured in terms of probability of misdetection $P_{\rm MD}$, defined as the probability that the \gls{ue} is not detected by the base station (i.e., the maximum \gls{snr} is below a threshold $\Gamma$) in an uplink scenario, or, vice versa, the base station is not detected by the \gls{ue} in a downlink scenario after a complete beam sweeping in all the available directions. 
The \textit{reactiveness} is the average time  to find and report the best beam pair for \gls{ia}, i.e., the time needed to perform the beam management procedures for \gls{ia} described in the previous section. 
Finally, the \textit{overhead} is the amount of time and frequency resources allocated to the framework with respect to the total amount of available resources. 

The simulations for the detection accuracy performance evaluation are based on realistic system design configurations where multiple \glspl{gnb} are deployed according to a Poisson Point Process.
The channel model is based on recent real-world measurements
at 28 GHz in New York City, to provide a realistic
assessment of mmWave micro and picocellular networks in
a dense urban deployment~\cite{akdeniz2014millimeter}.

\subsection{3GPP Framework Parameters}

In this section, we list the parameters that affect the performance of the measurement architectures, 
and provide insights on the impact of each parameter on the different~metrics.


\begin{figure}
	\centering
	\setlength\belowcaptionskip{-.8cm}
	\includegraphics[width=0.4\textwidth]{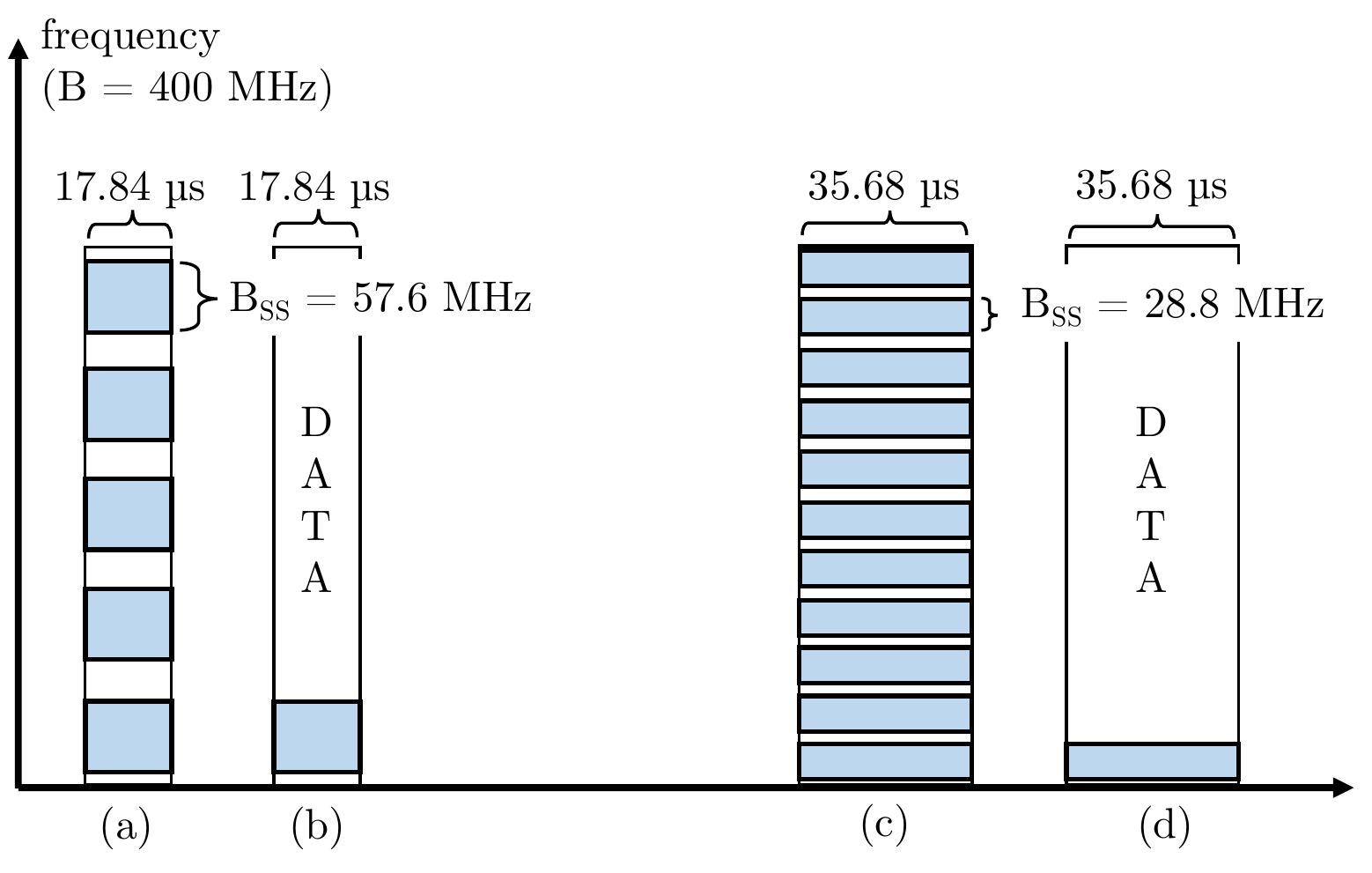}
	\caption{\gls{ss} block structure. Each blue rectangle is an \gls{ss} block (with 4 OFDM symbols).
	Cases (a) and (c) implement a \emph{frequency repetition} scheme (with $N_{rep}=5$ and $11$, respectively) while, for cases (b) and (d), a \emph{data} solution (i.e., $N_{rep}=1$) is preferred.}
	\label{fig:ssblock}
\end{figure}

As depicted in Fig. \ref{fig:ssblock}, we consider the \textit{frame structure} of 3GPP NR, with different subcarrier spacings $\Delta_f$.
For frequencies above 6 GHz, the subcarrier spacing $\Delta_f$ is $15\times2^n$ kHz, with $2 \le n \le 4$ (i.e., $\Delta_f = 60$, 120 or 240 kHz) and 14 \gls{ofdm} symbols per slot. The slot duration in $\mu$s is given by $  T_{\rm slot} = 1000/2^n$, 
while the duration of a symbol in $\mu$s is $  T_{\rm symb} = 71.42/2^n$~\cite{38211}.
Since the only subcarrier spacings considered for frequencies above 6 GHz are $\Delta_f = 120$ and  $240$ kHz, we will only consider these cases. Therefore, the slot duration is $125~\mu$s or $62.5~\mu$s, respectively. Moreover, since the maximum number of subcarriers allocated to the \gls{ss} blocks is 240, then the bandwidth allocated for the \gls{ss} blocks would be respectively 28.8 and 57.6 MHz. We consider a maximum channel bandwidth $B=400$ MHz per carrier.

Moreover, it is possible to configure the system to exploit \textit{frequency diversity}, $D$. Given that 240 subcarriers are allocated in frequency to an \gls{ss}, the remaining bandwidth in the symbols which contain an \gls{ss} block is $B - 240 \Delta_f$. Therefore, it is possible to adopt two different strategies: (i) \textit{data} (as represented in Figs. \ref{fig:ssblock}(b) and (d)), i.e., it is used for data transmission towards users which are in the same direction in which the \gls{ss} block is transmitted, or (ii) \textit{repetition} (as displayed in Figs. \ref{fig:ssblock}(a) and (c)), i.e., the information in the first 240 subcarriers is repeated in the remaining subcarriers to increase the robustness against noise and enhance the detection capabilities. The number of repetitions is therefore $N_{rep} = 1$ if frequency diversity is not used (i.e., $D=0$, and a single chunk of the available bandwidth is used for the \gls{ss} block), and $N_{rep} = 11$ or $N_{rep} = 5$ when repetition is used (i.e., $D=1$) with $\Delta_f = 120$~kHz or $\Delta_f = 240$~kHz, respectively. There is a guard interval
in frequency among the different repetitions of the \gls{ss} blocks, to provide a good tradeoff between frequency diversity and coherent combining~\cite{barati2015directional}.


We also consider different configurations of the \gls{ss} blocks and bursts. The maximum \textit{number $N_{\rm SS}$ of \gls{ss} blocks in a burst} for our frame structure and carrier frequencies is $L = 64$. We assume that, if $N_{\rm SS} < L$, the \gls{ss} blocks will be transmitted in the first $N_{\rm SS}$ opportunities. The actual maximum duration of an \gls{ss} burst is $D_{\max, \rm SS} = 2.5$~ms for $\Delta_f = 240$~kHz and $D_{\max, \rm SS} = 5$~ms for $\Delta_f = 120$~kHz. We will also investigate all the possible values for the \textit{\gls{ss} burst periodicity $T_{\rm SS}$}.

Another fundamental parameter is the \textit{array geometry}, i.e., the number of antenna elements $M$ at the \gls{gnb} and \gls{ue} and the number of directions that need to be covered, both in azimuth $N_{\theta}$ and in elevation $N_{\phi}$. At the \gls{gnb} we consider a single sector in a three-sector site, i.e., the azimuth $\theta$ varies from $-60$ to $60$ degrees, for a total of $\Delta_{\theta} = 120$ degrees, while the UE has a single array panel which covers the whole $\Delta_{\theta,\rm UE}=360^\circ$ angular space.
The elevation $\phi$ varies between $-30$ and $30$ degrees, for a total of $\Delta_{\phi} = 60$ degrees, and also includes a fixed mechanical tilt of the array pointing towards the ground. 
There exists a strong correlation among beamwidth, number of antenna elements and beamforming gain. The more antenna elements in the system, the narrower the beams, the higher the  gain that can be achieved by beamforming, and the more precise and directional the transmission. 
Thus, given the array geometry, we compute the beamwidth $\Delta_{\rm beam}$ at 3 dB of the main lobe of the beamforming vector, and then $N_{\theta} = \Delta_{\theta} / \Delta_{\rm beam}$ and $N_{\phi} = \Delta_{\phi} / \Delta_{\rm beam}$. The results are shown in Table~\ref{tab:BF}.

\begin{table}[t]
  \footnotesize
  \renewcommand{\arraystretch}{1}
  \centering
  \begin{tabular}{@{}llll@{}}
  \toprule
  $M$ & $\theta$ [deg]& $N_{\theta}$ gNB & $N_{\theta}$ UE  \\ \midrule
  4       & 60        & 2                          & 6                            \\
  16      & 26        & 5                          & 14                           \\
  64      & 13        & 10                         & 28          \\                \bottomrule
  \end{tabular}
  \caption{Relationship between $M$, $\theta$ and $N_{\theta}$, for the azimuth case~\cite{giordani2018tutorial}.}

  \label{tab:BF}
\end{table}

Additionally, different \textit{beamforming architectures}, i.e., analog, hybrid or digital, can be used both at the \gls{ue} and at the \gls{gnb}.
\textit{Analog beamforming} shapes the beam through a single \gls{rf} chain for all the antenna elements and therefore it is possible to transmit/receive in only one direction at any given time. This model saves power by using only a single pair of \glspl{adc}, but has small flexibility since the transceiver can only beamform in one direction.
\textit{Hybrid beamforming} uses $K_{\rm BF}$ \gls{rf} chains (with $1<K_{\rm BF}\leq M$), and thus is equivalent to $K_{\rm BF}$ parallel analog beams, as it enables the transceiver to transmit/receive in $K_{\rm BF}$ directions simultaneously. 
Nevertheless, when hybrid beamforming is used for transmission, the power available at each transmitting beam is the total node power constraint divided by $K_{\rm BF}$, thus potentially reducing the received power.
\textit{Digital beamforming} requires a separate \gls{rf} chain for each
antenna element and therefore  allows the processing of the received signals in the digital domain,  potentially enabling the transceiver to direct beams at infinitely many directions at the same time \cite{dutta2017fully}.
Although the digital transceiver is able to process an infinite number of received streams, only $M$ simultaneous and orthogonal beams can be handled without  significant inter-beam interference (i.e., through a zero-forcing beamforming structure \cite{yoo2006optimality}).  
For this reason, we limit the number of parallel  beams that can be generated to $M$. 
Furthermore, for energy-saving purposes, we implement a digital beamforming scheme only at the receiver side.
For the sake of completeness, we also consider an omnidirectional strategy at the \gls{ue}, i.e., without any beamforming gain but allowing the reception through the whole angular space at any given time.

Finally, the last parameter is the \textit{density of base stations }$\lambda_b$, expressed in \gls{gnb}/km$^2$.

\vspace{-0.1cm}
\section{Results and Discussion}
\vspace{-0.1cm}
\label{sec:results}
In this section, we present some simulation results aiming at 
(i) evaluating the performance of the presented \gls{ia} schemes in terms of detection accuracy (i.e., probability of misdetection), as reported in Sec.~\ref{sec:accuracy};
(ii) describing the analysis and the results related to the performance of the measurement frameworks for the reactiveness and the overhead, respectively in Sec.~\ref{sec:react} and Sec.~\ref{sec:overhead}.
Final considerations and remarks, aimed at providing guidelines to characterize the optimal IA configuration settings as a function of the system parameters, are contained in Sec.~\ref{sec:final_consid}.
\vspace{-0.1cm}
\subsection{Detection Accuracy Results}
\label{sec:accuracy}

\begin{figure}[t!]
  \centering
    \setlength\abovecaptionskip{0.1cm}
    \setlength\belowcaptionskip{-0.3cm}
  \setlength\fwidth{0.7\columnwidth}
  \setlength\fheight{0.3\columnwidth}
  \input{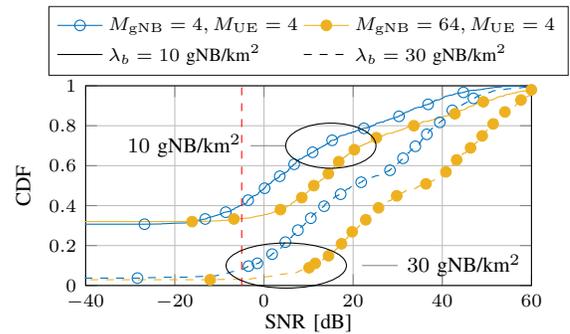}
  \caption{CDF of the SNR, for different antenna configurations. $\Delta_f=120$ kHz, $N_{rep}=1$. The red dashed line represents the SNR threshold $\Gamma=-5$ dB that has been considered throughout this work.}
    \label{fig:CDF_SNR}
\end{figure}

In Fig. \ref{fig:CDF_SNR} we plot the \gls{cdf} of the SNR between the mobile terminal and the gNB it is associated to, for different antenna configurations and considering two density values. Notice that the curves are not smooth because of the progressive transitions of the \gls{snr} among the different path loss regimes, i.e., \gls{los}, \gls{nlos} and outage.
We see that better detection accuracy performance can be achieved when densifying the network and when using larger arrays. 
In the first case, the endpoints are progressively closer, thus ensuring better signal quality and, in general, stronger received power. 
In the second case, narrower beams can be steered thus guaranteeing higher beamforming gains.
We also notice that, for good SNR regimes, the $ M_{\rm gNB} = 4, M_{\rm UE} = 4$ and $ M_{\rm gNB} = 64, M_{\rm UE} = 4$  configurations present good enough SNR values: in these regions, the channel conditions are sufficiently good to ensure satisfactory signal quality (and, consequently, acceptable misdetection) even when considering small antenna factors. 
Finally, the red line represents the SNR threshold $\Gamma=-5$ dB that we will consider in this~work.

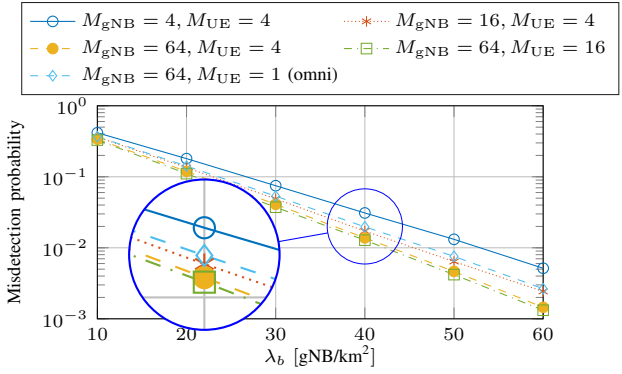
\begin{figure}[t]
  \centering
  \setlength\abovecaptionskip{0.1cm}
\setlength\belowcaptionskip{-.5cm}
  \setlength\fwidth{0.7\columnwidth}
  \setlength\fheight{0.32\columnwidth}
%
%
\usetikzlibrary{spy}
\definecolor{mycolor1}{rgb}{0.00000,0.44700,0.74100}%
\definecolor{mycolor2}{rgb}{0.85000,0.32500,0.09800}%
\definecolor{mycolor3}{rgb}{0.92900,0.69400,0.12500}%
\definecolor{mycolor4}{rgb}{0.49400,0.18400,0.55600}%
\definecolor{mycolor5}{rgb}{0.46600,0.67400,0.18800}%
\definecolor{mycolor6}{rgb}{0.30100,0.74500,0.93300}

\pgfplotsset{
tick label style={font=\scriptsize},
label style={font=\scriptsize},
legend  style={font=\scriptsize}
}

\begin{tikzpicture}[spy using outlines=
	{circle, magnification=2, connect spies}]
	
\begin{axis}[%
width=0.956\fwidth,
height=\fheight,
at={(0\fwidth,0\fheight)},
scale only axis,
xmin=10,
xmin=10,
xmax=60,
xlabel style={font=\color{white!15!black}},
xlabel={$\lambda_b$ $[\text{gNB/km}^\text{2}]$},
ylabel={Misdetection probability},
ymode=log,
ymin=0.001,
ymax=1,
xmajorgrids,
ymajorgrids,
xlabel shift=-5pt,
label style={font=\scriptsize},
legend columns=2,
legend style={font=\scriptsize, at={(0.5,1.05)},anchor=south,legend cell align=left, align=left, draw=white!15!black}
]
\addplot  [ color=mycolor1, mark=o, mark size=2.0pt, mark options={solid, mycolor1}]
  table[row sep=crcr]{%
10	0.41647\\
20	0.18026\\
30	0.07491\\
40	0.03094\\
50	0.01316\\
60	0.00515\\
};
\addlegendentry{$ M_{\rm gNB} = 4, M_{\rm UE} = 4$}

\addplot  [ color=mycolor2, densely dotted, mark=asterisk, mark size=2.0pt, mark options={solid, mycolor2}]
  table[row sep=crcr]{%
10	0.36243\\
20	0.13468\\
30	0.04889\\
40	0.01719\\
50	0.0064\\
60	0.00243\\
};
\addlegendentry{$ M_{\rm gNB} = 16, M_{\rm UE} = 4$}

\addplot [ color=mycolor3,mark=*, dashed, mark size=2.0pt, mark options={solid, mycolor3}]
  table[row sep=crcr]{%
10	0.33909\\
20	0.11887\\
30	0.04093\\
40	0.01379\\
50	0.00465\\
60	0.00145\\
};
\addlegendentry{$ M_{\rm gNB} = 64, M_{\rm UE} = 4$}

\addplot [color=mycolor5,dashdotted, mark=square, mark size=2.0pt, mark options={solid, mycolor5}]
  table[row sep=crcr]{%
10	0.32903\\
20	0.11071\\
30	0.0373\\
40	0.01281\\
50	0.00419\\
60	0.00132\\
};
\addlegendentry{$ M_{\rm gNB} = 64, M_{\rm UE} = 16$}

\addplot [color=mycolor6, mark=diamond, dashed, mark size=2.0pt, mark options={solid, mycolor6}]
  table[row sep=crcr]{%
10	0.36753\\
20	0.14157\\
30	0.05337\\
40	0.01975\\
50	0.00757\\
60	0.00265\\
};
\addlegendentry{$ M_{\rm gNB} = 64, M_{\rm UE} = 1 \text{ (omni)}$}

 \coordinate (spypoint) at (axis cs:40,0.02);
  \coordinate (magnifyglass) at (axis cs:22,0.008);
\end{axis}

\spy [blue, size=2cm] on (spypoint)
   in node[fill=white] at (magnifyglass);
\end{tikzpicture}%
  \caption{$P_{\rm MD}$ as a function of $\lambda_{b}$, for different antenna configurations.}
  \label{fig:P_err_N}
\end{figure}

\begin{figure}[t!]
  \centering    
    \setlength\abovecaptionskip{0.0cm}
  \setlength\belowcaptionskip{-.8cm}
  \setlength\fwidth{0.7\columnwidth}
  \setlength\fheight{0.32\columnwidth}
%
%
\usetikzlibrary{spy}
\definecolor{mycolor1}{rgb}{0.00000,0.44700,0.74100}%
\definecolor{mycolor2}{rgb}{0.85000,0.32500,0.09800}%
\definecolor{mycolor3}{rgb}{0.92900,0.69400,0.12500}%
\definecolor{mycolor4}{rgb}{0.49400,0.18400,0.55600}%
\definecolor{mycolor5}{rgb}{0.46600,0.67400,0.18800}%
\definecolor{mycolor6}{rgb}{0.30100,0.74500,0.93300}

\pgfplotsset{
tick label style={font=\scriptsize},
label style={font=\scriptsize},
legend  style={font=\scriptsize}
}

\begin{tikzpicture}[spy using outlines=
	{circle, magnification=5, connect spies}]

\begin{axis}[%
width=0.956\fwidth,
height=\fheight,
at={(0\fwidth,0\fheight)},
scale only axis,
xmin=10,
xmax=60,
xlabel style={font=\color{white!15!black}},
xlabel={$\lambda_b$ $[\text{BS/km}^\text{2}]$},
ymode=log,
ymin=0.001,
ymax=1,
yminorticks=true,
xmajorgrids,
ymajorgrids,
ylabel style={font=\color{white!15!black}},
ylabel={Misdetection probability},
axis background/.style={fill=white},
legend columns=2,
xlabel shift=-5pt,
label style={font=\scriptsize},
legend style={font=\scriptsize, at={(0.5,1.05)},anchor=south,legend cell align=left, align=left, draw=white!15!black}
]

\addplot [color=black, mark=x,mark size=4pt, mark options={solid, fill=red, red}]
  table[row sep=crcr]{%
10	0.41647\\\\
};
\addlegendentry{ $\Delta_f=120$ kHz, $D=0$}

\addplot [color=black, mark=square*,mark size=2pt, mark options={solid, fill=black, black}]
  table[row sep=crcr]{%
10	0.42265\\
};
\addlegendentry{ $\Delta_f=120$ kHz, $D=1$}

\addplot [color=black, mark=o,mark size=2pt, mark options={solid, fill=black, black}]
  table[row sep=crcr]{%
10	0.44092\\
};
\addlegendentry{ $\Delta_f=240$ kHz, $D=1$}

\addplot [color=black, mark=triangle*,mark size=3pt, mark options={solid, fill=black, black}]
  table[row sep=crcr]{%
10	0.46164\\
};
\addlegendentry{ $\Delta_f=240$ kHz, $D=0$}

\addplot  [ forget plot, color=mycolor1, mark=square*, mark size=2.0pt, mark options={solid, mycolor1}]
  table[row sep=crcr]{%
10	0.3772\\
20	0.14623\\
30	0.05532\\
40	0.02132\\
50	0.00812\\
60	0.00319\\
};
\addlegendentry{4x4, Spacing = 120 kHz, freq. diversity}

\addplot[forget plot,dashed,color=mycolor1, mark=x, mark size=2.0pt, mark options={solid, red}]
  table[row sep=crcr]{%
10	0.41709\\
20	0.17705\\
30	0.07494\\
40	0.03177\\
50	0.01358\\
60	0.00569\\
};
\addlegendentry{4x4, Spacing = 120 kHz, NO freq. diversity}

\addplot [forget plot,color=mycolor1, mark=o, mark size=4.0pt, mark options={solid}]
  table[row sep=crcr]{%
10	0.41954\\
20	0.17947\\
30	0.07494\\
40	0.03228\\
50	0.01345\\
60	0.0056\\
};
\addlegendentry{4x4, Spacing = 240 kHz, freq. diversity}

\addplot [forget plot,color=mycolor1, mark=triangle*, mark size=2.0pt, mark options={solid, mycolor1}]
  table[row sep=crcr]{%
10	0.45907\\
20	0.21563\\
30	0.09824\\
40	0.04558\\
50	0.02133\\
60	0.00967\\
};
\addlegendentry{4x4, Spacing = 240 kHz, NO freq. diversity}

\addplot[ forget plot, color=mycolor3, mark=square*, mark size=2.0pt, mark options={solid, mycolor3}]
  table[row sep=crcr]{%
10	0.33103\\
20	0.11234\\
30	0.03773\\
40	0.01269\\
50	0.00406\\
60	0.00159\\
};
\addlegendentry{64x4, Spacing = 120 kHz, freq. diversity}

\addplot [forget plot,dashed,color=mycolor3, mark=x, mark size=2.0pt, mark options={solid, red}]
  table[row sep=crcr]{%
10	0.33887\\
20	0.11752\\
30	0.04032\\
40	0.01402\\
50	0.0046\\
60	0.00183\\
};
\addlegendentry{64x4, Spacing = 120 kHz, NO freq. diversity}

\addplot [forget plot,color=mycolor3, mark=o, mark size=4.0pt, mark options={solid}]
  table[row sep=crcr]{%
10	0.33715\\
20	0.11816\\
30	0.03913\\
40	0.01359\\
50	0.0047\\
60	0.00153\\
};
\addlegendentry{64x4, Spacing = 240 kHz, freq. diversity}

\addplot [forget plot,color=mycolor3, mark=triangle*, mark size=2.0pt, mark options={solid, mycolor3}]
  table[row sep=crcr]{%
10	0.34845\\
20	0.12596\\
30	0.04325\\
40	0.01556\\
50	0.00561\\
60	0.00189\\
};
\addlegendentry{64x4, Spacing = 240 kHz, NO freq. diversity}

\node[coordinate] (A) at (axis cs:35,0.049) {};                       
\node[coordinate,pin=above:{\footnotesize $M_{\rm gNB} = 4,  M_{\rm UE} = 4$}] at (axis cs:35,0.085){};  

\node[coordinate] (B) at (axis cs:45,0.008) {};                       
\node[coordinate,pin=left:{\footnotesize $M_{\rm gNB} = 64,  M_{\rm UE} = 4$}] at (axis cs:42,0.008){};  

\end{axis}
\draw[mycolor1] (A) ellipse (0.3 and 0.45);                                
\draw[mycolor3] (B) ellipse (0.3 and 0.2);                                
\end{tikzpicture}%
  \caption{$P_{\rm MD}$ as a function of $\lambda_{b}$, for different subcarrier spacings $\Delta_f$ and repetition strategies and for different antenna configurations.  $ M_{\rm gNB} = 4, M_{\rm UE} = 4$, $\Gamma=-5$ dB.}
  \label{fig:D_4x4}
\end{figure}
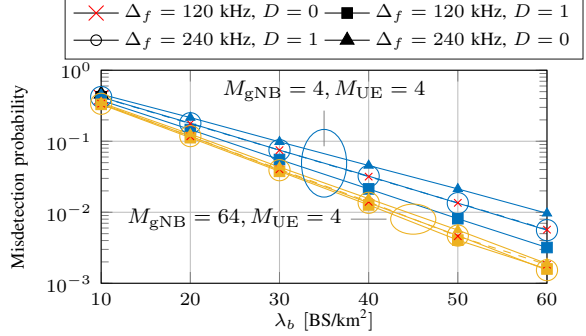

Analogous considerations can be deduced from Fig. \ref{fig:P_err_N} which illustrates how the misdetection probability monotonically decreases when the gNB density $\lambda_b$ progressively increases or when the transceiver is equipped with a larger number of antenna elements, since more focused beams can be generated in this case. 
Moreover, we notice that the beamforming strategy in which the UE transmits or receives omnidirectionally, although guaranteeing fast access operations, does not ensure accurate \gls{ia} performance  and leads to degraded detection capabilities. More specifically, the gap with a fully directional architecture (e.g., $ M_{\rm gNB} = 64, M_{\rm UE} = 16$) is quite remarkable for very dense scenarios, and increases as the \gls{gnb} density increases. For example, the configuration with 16 antennas (i.e., $M_{\rm UE} =16$) and that with a single omnidirectional antenna at the \gls{ue} reach the same $P_{\rm MD}$, but at different values of \gls{gnb} density $\lambda_b$, respectively 30 and 35 $[\text{gNB/km}^\text{2}]$: the omnidirectional configuration requires a higher density (i.e., 5 $[\text{gNB/km}^\text{2}]$ more) to compensate for the smaller beamforming gain.


Finally, Fig. \ref{fig:D_4x4} reports the misdetection probability related to $\lambda_b$, for different subcarrier spacings $\Delta_f$ and repetition strategies $D$.
First, we see that, if no repetitions are used (i.e., $D=0$), lower detection accuracy performance is associated with the $\Delta_f=240$ kHz configuration, due to the resulting larger impact of the thermal noise and the consequent SNR degradation. 
Furthermore, the detection efficiency can be enhanced by repeating the SS block information embedded in the first 240 subcarriers in the remaining subcarriers (i.e., $D=1$), to increase the robustness of the communication and mitigate the effect of the noise in the detection process. 
In fact, if a frequency diversity approach is preferred, the UE (in the DL measurement technique) or the gNB (in the UL measurement technique) has $N_{rep}>1$ attempts to properly collect the synchronization signals exchanged during the beam sweeping phase, compared to the single opportunity the nodes would have had if they had not implemented any repetition strategy.
We also observe that the  $\Delta_f=120$ kHz with no frequency diversity configuration and the $\Delta_f=240$ kHz scheme with $N_{rep}=5$ produce the same detection accuracy results, thus showing that increasing the subcarrier spacing and increasing the number of repetitions of the SS block information in multiple frequency subbands have almost the same effect in terms of misdetection capabilities.
Finally, we observe that the impact of the frequency diversity $D$ and the subcarrier spacing $\Delta_f$ is less significant when increasing the array factor, as can be seen from the reduced gap between the curves plotted in Fig. \ref{fig:D_4x4} for the $M_{\rm gNB} = 4, M_{\rm UE} = 4$ and $M_{\rm gNB} = 64, M_{\rm UE} = 4$ configurations.
The reason is that, when considering larger arrays, even the configuration with $\Delta_f = 240$~kHz, with no repetitions, has an average \gls{snr} which is high enough to reach small misdetection probability values.

\subsection{Reactiveness Results}
\label{sec:react}
For \gls{ia}, reactiveness is defined as the delay required to perform a full iterative search in all the possible combinations of the directions. 
The \gls{gnb} and the \gls{ue} need to scan respectively $N_{\theta,\rm gNB}N_{\phi,\rm gNB}$ and $N_{\theta,\rm UE}N_{\phi,\rm UE}$ directions to cover the whole horizontal and vertical space. Moreover, they can transmit or receive respectively $K_{\rm BF, gNB}$ and $K_{\rm BF, UE}$ beams simultaneously. Notice that, as mentioned in Sec. \ref{sec:system_model},  for digital and omnidirectional architectures $K_{\rm BF} = \min\{N_{\theta}N_{\phi}, M\}$,
for hybrid $K_{\rm BF} = \min\{N_{\theta}N_{\phi},M\}/\nu$, where $\nu$ is a factor that limits the number of directions in which it is possible to transmit or receive at the same time, and for analog $K_{\rm BF} = 1$ \cite{sun2014MIMO}.
Then the total number of \gls{ss} blocks needed is
\begin{equation}
	S_D = \left\lceil\frac{N_{\theta,\rm gNB}N_{\phi,\rm gNB}}{K_{\rm BF, gNB}}\right\rceil\left\lceil\frac{N_{\theta,\rm UE}N_{\phi,\rm UE}}{K_{\rm BF, UE}}\right\rceil.
\label{eq:S_D}
\end{equation}

\begin{figure}[t!]
  \centering
  \setlength\belowcaptionskip{-.5cm}
  \begin{subfigure}[t!]{\columnwidth}
    \flushright
    \setlength\belowcaptionskip{0.1cm}
    \setlength\abovecaptionskip{0cm}
    \setlength\fwidth{0.7\columnwidth}
    \setlength\fheight{0.30\columnwidth}
%
%
\definecolor{mycolor1}{rgb}{0.00000,0.44700,0.74100}%
\definecolor{mycolor2}{rgb}{0.85000,0.32500,0.09800}%
\definecolor{mycolor3}{rgb}{0.92900,0.69400,0.12500}%
\definecolor{mycolor4}{rgb}{0.49400,0.18400,0.55600}%
\definecolor{mycolor5}{rgb}{0.46600,0.67400,0.18800}%
\definecolor{mycolor6}{rgb}{0.30100,0.74500,0.93300}%
\pgfplotsset{scaled y ticks=false}
\begin{tikzpicture}
\pgfplotsset{every tick label/.append style={font=\scriptsize}}

\begin{axis}[%
width=0.956\fwidth,
height=\fheight,
at={(0\fwidth,0\fheight)},
scale only axis,
xmin=7,
xmax=65,
xtick=data,
xlabel style={font=\scriptsize\color{white!15!black}},
xlabel={$N_{\rm SS}$},
ymin=0,
ymax=5300,
ylabel style={font=\scriptsize\color{white!15!black}},
ylabel={$T_{\rm IA}$ [ms]},
axis background/.style={fill=white},
xmajorgrids,
ymajorgrids,
xlabel shift=-5pt,
legend columns=2,
legend style={at={(0.5, 1.05)},anchor=south,font=\scriptsize,legend cell align=left, align=left, draw=white!15!black}
]
\addplot [color=mycolor1, line width=1.2pt, mark=o, mark options={solid, mycolor1}]
  table[row sep=crcr]{%
8	20.11608125\\
16	0.36608125\\
32	0.36608125\\
64	0.36608125\\
};
\addlegendentry{$ M_{\rm gNB} = 4, M_{\rm UE} = 4$}

\addplot [color=mycolor2,  densely dotted, line width=1.2pt, mark=asterisk, mark options={solid, mycolor2}]
  table[row sep=crcr]{%
8	220.05358125\\
16	100.30358125\\
32	40.80358125\\
64	20.80358125\\
};
\addlegendentry{$ M_{\rm gNB} = 16, M_{\rm UE} = 4$}

\addplot [color=mycolor3, dashed, line width=1.2pt, mark=*, mark options={solid, mycolor3}]
  table[row sep=crcr]{%
8	740.11608125\\
16	360.36608125\\
32	180.36608125\\
64	81.36608125\\
};
\addlegendentry{$ M_{\rm gNB} = 64, M_{\rm UE} = 4$}

\addplot [color=mycolor4, dotted, line width=1.2pt, mark=x, mark options={solid, mycolor4}]
  table[row sep=crcr]{%
8	1560.17858125\\
16	780.17858125\\
32	380.67858125\\
64	181.67858125\\
};
\addlegendentry{$ M_{\rm gNB} = 16, M_{\rm UE} = 16$}

\addplot [color=mycolor5, dashdotted, line width=1.2pt, mark=square, mark options={solid, mycolor5}]
  table[row sep=crcr]{%
8	5240.11608125\\
16	2620.11608125\\
32	1300.61608125\\
64	641.61608125\\
};
\addlegendentry{$ M_{\rm gNB} = 64, M_{\rm UE} = 16$}

\addplot [color=mycolor6, dashed, line width=1.2pt, mark=diamond, mark options={solid, mycolor6}]
  table[row sep=crcr]{%
8	120.05358125\\
16	60.05358125\\
32	20.55358125\\
64	1.55358125\\
};
\addlegendentry{$ M_{\rm gNB} = 64, M_{\rm UE} = 1$ $(\mbox{omni})$}

\end{axis}
\end{tikzpicture}%
    \caption{\gls{gnb} Analog, UE Analog}
    \label{fig:nsseNBAnUeAn}
  \end{subfigure}
  \begin{subfigure}[t!]{\columnwidth}
  \centering
    \setlength\belowcaptionskip{0.1cm}
    \setlength\abovecaptionskip{0cm}
    \setlength\fwidth{0.7\columnwidth}
    \setlength\fheight{0.30\columnwidth}
%
%
\definecolor{mycolor1}{rgb}{0.00000,0.44700,0.74100}%
\definecolor{mycolor2}{rgb}{0.85000,0.32500,0.09800}%
\definecolor{mycolor3}{rgb}{0.92900,0.69400,0.12500}%
\definecolor{mycolor4}{rgb}{0.49400,0.18400,0.55600}%
\definecolor{mycolor5}{rgb}{0.46600,0.67400,0.18800}%
\definecolor{mycolor6}{rgb}{0.30100,0.74500,0.93300}%
\pgfplotsset{scaled y ticks=false}
\begin{tikzpicture}
\pgfplotsset{every tick label/.append style={font=\scriptsize}}

\begin{axis}[%
width=0.956\fwidth,
height=\fheight,
at={(0\fwidth,0\fheight)},
scale only axis,
xmin=7,
xmax=65,
xtick=data,
xlabel style={font=\scriptsize\color{white!15!black}},
xlabel={$N_{\rm SS}$},
ymin=0,
ymax=365,
ylabel style={font=\scriptsize\color{white!15!black}},
ylabel={$T_{\rm IA}$ [ms]},
axis background/.style={fill=white},
xmajorgrids,
ymajorgrids,
xlabel shift=-5pt,
legend style={font=\scriptsize,legend cell align=left, align=left, draw=white!15!black,at={(0.99,0.6)},anchor=south east}
]
\addplot [color=mycolor1, line width=1.2pt, mark=o, mark options={solid, mycolor1}]
  table[row sep=crcr]{%
8 0.11608125\\
16  0.11608125\\
32  0.11608125\\
64  0.11608125\\
};
\addlegendentry{$ M_{\rm gNB} = 4, M_{\rm UE} = 4$}

\addplot [color=mycolor2, densely dotted, line width=1.2pt, mark=asterisk, mark options={solid, mycolor2}]
  table[row sep=crcr]{%
8 60.17858125\\
16  20.42858125\\
32  0.92858125\\
64  0.92858125\\
};
\addlegendentry{$ M_{\rm gNB} = 16, M_{\rm UE} = 4$}

\addplot [color=mycolor3, dashed, line width=1.2pt, mark=*, mark options={solid, mycolor3}]
  table[row sep=crcr]{%
8 240.11608125\\
16  120.11608125\\
32  60.11608125\\
64  21.11608125\\
};
\addlegendentry{$ M_{\rm gNB} = 64, M_{\rm UE} = 4$}

\addplot [color=mycolor4, dotted, line width=1.2pt, mark=x, mark options={solid, mycolor4}]
  table[row sep=crcr]{%
8 100.15175625\\
16  40.40175625\\
32  20.40175625\\
64  1.40175625\\
};
\addlegendentry{$ M_{\rm gNB} = 16, M_{\rm UE} = 16$}

\addplot [color=mycolor5, dashdotted, line width=1.2pt, mark=square, mark options={solid, mycolor5}]
  table[row sep=crcr]{%
8 360.17858125\\
16  180.17858125\\
32  80.67858125\\
64  40.67858125\\
};
\addlegendentry{$ M_{\rm gNB} = 64, M_{\rm UE} = 16$}

\addplot [color=mycolor6, dashed, line width=1.2pt, mark=diamond, mark options={solid, mycolor6}]
  table[row sep=crcr]{%
8 120.05358125\\
16  60.05358125\\
32  20.55358125\\
64  1.55358125\\
};
\addlegendentry{$ M_{\rm gNB} = 64, M_{\rm UE} = 1$ $(\mbox{omni})$}

\coordinate (pt) at (axis cs:32,0);

\legend{};

\end{axis}

\node[pin=3:{%
    \begin{tikzpicture}[trim axis left,trim axis right]
    \pgfplotsset{every tick label/.append style={font=\tiny}}
    \begin{axis}[
    scale only axis,
bar shift auto,
      xmin=30,xmax=34,
      ymin=0,ymax=1.2,
width=0.2\fwidth,
height=0.4\fheight,
axis x line*=bottom,
axis y line*=left, 
xtick=data,
axis background/.style={fill=white},
    ]
\addplot [color=mycolor1, line width=1.2pt, mark=o, mark options={solid, mycolor1}]
  table[row sep=crcr]{%
8 0.11608125\\
16  0.11608125\\
32  0.11608125\\
64  0.11608125\\
};

\addplot [color=mycolor2, densely dotted, line width=1.2pt, mark=asterisk, mark options={solid, mycolor2}]
  table[row sep=crcr]{%
8 60.17858125\\
16  20.42858125\\
32  0.92858125\\
64  0.92858125\\
};

\addplot [color=mycolor3, dashed, line width=1.2pt, mark=*, mark options={solid, mycolor3}]
  table[row sep=crcr]{%
8 240.11608125\\
16  120.11608125\\
32  60.11608125\\
64  21.11608125\\
};

\addplot [color=mycolor4, dotted, line width=1.2pt, mark=x, mark options={solid, mycolor4}]
  table[row sep=crcr]{%
8 100.15175625\\
16  40.40175625\\
32  20.40175625\\
64  1.40175625\\
};

\addplot [color=mycolor5, dashdotted, line width=1.2pt, mark=square, mark options={solid, mycolor5}]
  table[row sep=crcr]{%
8 360.17858125\\
16  180.17858125\\
32  80.67858125\\
64  40.67858125\\
};

\addplot [color=mycolor6, dashed, line width=1.2pt, mark=diamond, mark options={solid, mycolor6}]
  table[row sep=crcr]{%
8 120.05358125\\
16  60.05358125\\
32  20.55358125\\
64  1.55358125\\
};

    \end{axis}
    \end{tikzpicture}%
}] at (pt) {};

\end{tikzpicture}%
    \caption{\gls{gnb} Analog, UE Digital (\gls{dl}-based configuration)}
    \label{fig:nsseNBAnUeDig}
  \end{subfigure}
  \begin{subfigure}[t!]{\columnwidth}
  \centering
    \setlength\belowcaptionskip{0.1cm}
    \setlength\abovecaptionskip{0cm}
    \setlength\fwidth{0.7\columnwidth}
    \setlength\fheight{0.30\columnwidth}
%
%
\definecolor{mycolor1}{rgb}{0.00000,0.44700,0.74100}%
\definecolor{mycolor2}{rgb}{0.85000,0.32500,0.09800}%
\definecolor{mycolor3}{rgb}{0.92900,0.69400,0.12500}%
\definecolor{mycolor4}{rgb}{0.49400,0.18400,0.55600}%
\definecolor{mycolor5}{rgb}{0.46600,0.67400,0.18800}%
\definecolor{mycolor6}{rgb}{0.30100,0.74500,0.93300}%
\pgfplotsset{scaled y ticks=false}
\begin{tikzpicture}
\pgfplotsset{every tick label/.append style={font=\scriptsize}}

\begin{axis}[%
width=0.956\fwidth,
height=\fheight,
at={(0\fwidth,0\fheight)},
scale only axis,
xmin=7,
xmax=65,
xtick=data,
xlabel style={font=\scriptsize\color{white!15!black}},
xlabel={$N_{\rm SS}$},
ymin=0,
ymax=110,
ylabel style={font=\scriptsize\color{white!15!black}},
ylabel={$T_{\rm IA}$ [ms]},
axis background/.style={fill=white},
xmajorgrids,
ymajorgrids,
xlabel shift=-5pt,
legend style={font=\scriptsize,legend cell align=left, align=left, draw=white!15!black, at={(0.99,0.6)},anchor=south east}
]
\addplot [color=mycolor1, line width=1.2pt, mark=o, mark options={solid, mycolor1}]
  table[row sep=crcr]{%
8 0.17858125\\
16  0.17858125\\
32  0.17858125\\
64  0.17858125\\
};
\addlegendentry{$ M_{\rm gNB} = 4, M_{\rm UE} = 4$}

\addplot [color=mycolor2, densely dotted, line width=1.2pt, mark=asterisk, mark options={solid, mycolor2}]
  table[row sep=crcr]{%
8 0.17858125\\
16  0.17858125\\
32  0.17858125\\
64  0.17858125\\
};
\addlegendentry{$ M_{\rm gNB} = 16, M_{\rm UE} = 4$}

\addplot [color=mycolor3, dashed, line width=1.2pt, mark=*, mark options={solid, mycolor3}]
  table[row sep=crcr]{%
8 0.17858125\\
16  0.17858125\\
32  0.17858125\\
64  0.17858125\\
};
\addlegendentry{$ M_{\rm gNB} = 64, M_{\rm UE} = 4$}

\addplot [color=mycolor4, dotted, line width=1.2pt, mark=x, mark options={solid, mycolor4}]
  table[row sep=crcr]{%
8 100.05358125\\
16  40.30358125\\
32  20.30358125\\
64  1.30358125\\
};
\addlegendentry{$ M_{\rm gNB} = 16, M_{\rm UE} = 16$}

\addplot [color=mycolor5, dashdotted, line width=1.2pt, mark=square, mark options={solid, mycolor5}]
  table[row sep=crcr]{%
8 100.05358125\\
16  40.30358125\\
32  20.30358125\\
64  1.30358125\\
};
\addlegendentry{$ M_{\rm gNB} = 64, M_{\rm UE} = 16$}

\addplot [color=mycolor6, dashed, line width=1.2pt, mark=diamond, mark options={solid, mycolor6}]
  table[row sep=crcr]{%
8 0.02675625\\
16  0.02675625\\
32  0.02675625\\
64  0.02675625\\
};
\addlegendentry{$ M_{\rm gNB} = 64, M_{\rm UE} = 1$ $(\mbox{omni})$}

\coordinate (pt) at (axis cs:32,0);

\legend{};

\end{axis}

\node[pin=3:{%
    \begin{tikzpicture}[trim axis left,trim axis right]
    \pgfplotsset{every tick label/.append style={font=\tiny}}
    \begin{axis}[
    scale only axis,
bar shift auto,
      xmin=30,xmax=34,
      ymin=0,ymax=1.2,
width=0.2\fwidth,
height=0.4\fheight,
axis x line*=bottom,
axis y line*=left, 
xtick=data,
axis background/.style={fill=white},
    ]
\addplot [color=mycolor1, line width=1.2pt, mark=o, mark options={solid, mycolor1}, forget plot]
  table[row sep=crcr]{%
8	0.17858125\\
16	0.17858125\\
32	0.17858125\\
64	0.17858125\\
};

\addplot [color=mycolor2, densely dotted, line width=1.2pt, mark=asterisk, mark options={solid, mycolor2}, forget plot]
  table[row sep=crcr]{%
8	0.17858125\\
16	0.17858125\\
32	0.17858125\\
64	0.17858125\\
};

\addplot [color=mycolor3, dashed, line width=1.2pt, mark=*, mark options={solid, mycolor3}, forget plot]
  table[row sep=crcr]{%
8	0.17858125\\
16	0.17858125\\
32	0.17858125\\
64	0.17858125\\
};

\addplot [color=mycolor4, dotted,  line width=1.2pt, mark=x, mark options={solid, mycolor4}, forget plot]
 table[row sep=crcr]{%
8 100.05358125\\
16  40.30358125\\
32  20.30358125\\
64  1.30358125\\
};

\addplot [color=mycolor5, dashdotted, line width=1.2pt, mark=square, mark options={solid, mycolor5}, forget plot]
  table[row sep=crcr]{%
8 100.05358125\\
16  40.30358125\\
32  20.30358125\\
64  1.30358125\\
};

\addplot [color=mycolor6, dashed, line width=1.2pt, mark=diamond, mark options={solid, mycolor6}, forget plot]
  table[row sep=crcr]{%
8	0.02675625\\
16	0.02675625\\
32	0.02675625\\
64	0.02675625\\
};

    \end{axis}
    \end{tikzpicture}%
}] at (pt) {};

\end{tikzpicture}%
    \caption{\gls{gnb} Digital, UE Analog (\gls{ul}-based configuration)}
    \label{fig:nsseNBDigUeAn}
  \end{subfigure}
  \caption{$T_{\rm IA}$ as a function of $N_{\rm SS}$ with $T_{\rm SS} = 20$~ms.}
  \label{fig:nss}
\end{figure}
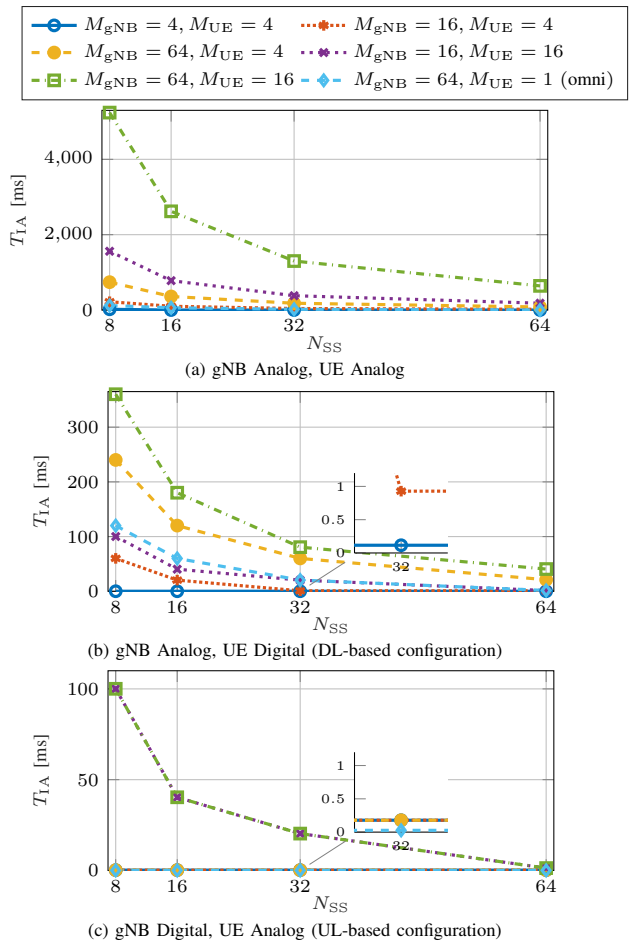

Given that there are $N_{\rm SS}$ blocks in a burst, the total delay from the beginning of an \gls{ss} burst transmission in a \gls{gnb} to the completion of the sweep in all the possible directions~is
\begin{equation}\label{eq:reactIa}
	T_{\rm IA} = T_{\rm SS}\left(\left\lceil\frac{S_D}{N_{SS}}\right\rceil - 1\right) + T_{last},
\end{equation}
where $T_{last}$ is the time required to transmit the remaining \gls{ss} blocks in the last burst (notice that there may be just one burst, in which case the first term in Eq.~\eqref{eq:reactIa} would be 0). This term depends on the subcarrier spacing and on the number of remaining \gls{ss} blocks which is given by 
\begin{equation}
	N_{\rm SS, left} = S_D - N_{\rm SS}\left(\left\lceil\frac{S_D}{N_{\rm SS}}\right\rceil - 1\right).
\end{equation}
Then, $T_{last}$ is
\begin{equation}
T_{last} = 
	\begin{cases}
		\frac{N_{\rm SS, left}}{2}T_{slot} - 2T_{symb} & \mbox{ if }  N_{\rm SS, left}\bmod 2   = 0\\
		\left\lfloor\frac{N_{\rm SS, left}}{2}\right\rfloor T_{slot} + 6T_{symb} & \mbox{ otherwise,}
	\end{cases}
	\label{eq:T_last}
\end{equation}

The two different options account for an even or odd remaining number of \gls{ss} blocks. In the first case, the \gls{ss} blocks are sent in $N_{\rm SS, left}/2$ slots, with total duration $ T_{slot} N_{\rm SS, left}/2$, but the last one is actually received in the 12\textit{th} symbol of the last slot, i.e., 2 symbols before the end of that slot, given the positions of the \gls{ss} blocks in each slot described in~\cite{38211}. If instead $N_{\rm SS, left}$ is odd, six symbols of slot $\lfloor N_{\rm SS, left} / 2 \rfloor + 1$ are also~used.

A selection of results is presented in the next paragraphs. 
In Fig.~\ref{fig:nss} we consider first the impact of the number of \gls{ss} blocks in a burst, with a fixed \gls{ss} burst periodicity $T_{\rm SS} = 20$~ms and  for different beamforming strategies and antenna configurations. 
In particular in Fig.~\ref{fig:nsseNBAnUeAn}, in which both the \gls{ue} and the \gls{gnb} use analog beamforming,  the \gls{ia} delay heavily depends on the number of antennas at the transceivers since all the available directions must be scanned one by one. 
It may take from 0.6 s (with $N_{\rm SS} = 64$) up to 5.2 s (with $N_{\rm SS} = 8$) to transmit and receive all the possible beams, which makes the scheme infeasible for practical usage.
A reduction in the sweeping time can be achieved either by using an omnidirectional antenna at the \gls{ue} or by decreasing the number of directions to be scanned both at the \gls{ue} and at the \gls{gnb}.
In this case, the only configurations that manage to complete a scan in a single \gls{ss} burst are those with 4 antennas at both sides and $N_{\rm SS} \ge 16$, or that with $M_{\rm gNB} = 64$, an omnidirectional \gls{ue} and $N_{\rm SS} = 64$.
Another option is the usage of digital beamforming at the \gls{ue} in a downlink-based scheme (as displayed in Fig. \ref{fig:nsseNBAnUeDig}), or at the \gls{gnb} in an uplink-based one (as shown in Fig.~\ref{fig:nsseNBDigUeAn}). 
This increases the number of configurations able to complete a sweep in an \gls{ss} block, even with a large number of antennas at the \gls{gnb} and the \gls{ue}. 
In particular, Fig.~\ref{fig:nsseNBDigUeAn} shows the performance of an uplink-based scheme, in which the \glspl{srs} are sent in the same time and frequency resource in which the \gls{ss} blocks would be sent, and the \gls{gnb} uses digital beamforming. It can be seen that there is a gain in performance for most of the configurations, because the \gls{gnb} has to sweep more directions than the \gls{ue} (since it uses narrower beams), thus using digital beamforming at the \gls{gnb} side makes it possible to reduce $T_{\rm IA}$ even more than when it is used at the \gls{ue} side.

\begin{figure}[t!]
  \centering
  \setlength\belowcaptionskip{-.6cm}
    \setlength\fwidth{0.7\columnwidth}
    \setlength\fheight{0.3\columnwidth}
%
%
\definecolor{mycolor1}{rgb}{0.00000,0.44700,0.74100}%
\definecolor{mycolor2}{rgb}{0.85000,0.32500,0.09800}%
\definecolor{mycolor3}{rgb}{0.92900,0.69400,0.12500}%
\definecolor{mycolor4}{rgb}{0.49400,0.18400,0.55600}%
\definecolor{mycolor5}{rgb}{0.46600,0.67400,0.18800}%
\definecolor{mycolor6}{rgb}{0.30100,0.74500,0.93300}%
\pgfplotsset{scaled y ticks=false}
\begin{tikzpicture}
\pgfplotsset{every tick label/.append style={font=\scriptsize}}
\begin{axis}[%
width=0.956\fwidth,
height=\fheight,
at={(0\fwidth,0\fheight)},
scale only axis,
xmin=4,
xmax=161,
xlabel style={font=\scriptsize\color{white!15!black}},
xlabel={$T_{\rm SS}$ [ms]},
ymin=0,
ymax=650,
ylabel style={font=\scriptsize\color{white!15!black}},
ylabel={$T_{\rm IA}$ [ms]},
axis background/.style={fill=white},
xmajorgrids,
ymajorgrids,
legend style={font=\scriptsize,at={(0.01, 0.55)},anchor=south west,legend cell align=left, align=left, draw=white!15!black}
]
\addplot [color=mycolor1, line width=1.2pt, mark=o, mark options={solid, mycolor1}]
  table[row sep=crcr]{%
5	0.17858125\\
10	0.17858125\\
20	0.17858125\\
40	0.17858125\\
80	0.17858125\\
160	0.17858125\\
};
\addlegendentry{$ M_{\rm gNB} = 4, M_{\rm UE} = 4$}

\addplot [color=mycolor2, densely dotted, line width=1.2pt, mark=asterisk, mark options={solid, mycolor2}]
  table[row sep=crcr]{%
5	1.40175625\\
10	1.40175625\\
20	1.40175625\\
40	1.40175625\\
80	1.40175625\\
160	1.40175625\\
};
\addlegendentry{$ M_{\rm gNB} = 16, M_{\rm UE} = 4$}

\addplot [color=mycolor3, dashed, line width=1.2pt, mark=*, mark options={solid, mycolor3}]
  table[row sep=crcr]{%
5	10.67858125\\
10	20.67858125\\
20	40.67858125\\
40	80.67858125\\
80	160.67858125\\
160	320.67858125\\
};
\addlegendentry{$ M_{\rm gNB} = 64, M_{\rm UE} = 4$}

\addplot [color=mycolor4, dotted, line width=1.2pt, mark=x, mark options={solid, mycolor4}]
  table[row sep=crcr]{%
5	5.80358125\\
10	10.80358125\\
20	20.80358125\\
40	40.80358125\\
80	80.80358125\\
160	160.80358125\\
};
\addlegendentry{$ M_{\rm gNB} = 16, M_{\rm UE} = 16$}

\addplot [color=mycolor5, dashdotted, line width=1.2pt, mark=square, mark options={solid, mycolor5}]
  table[row sep=crcr]{%
5	21.36608125\\
10	41.36608125\\
20	81.36608125\\
40	161.36608125\\
80	321.36608125\\
160	641.36608125\\
};
\addlegendentry{$ M_{\rm gNB} = 64, M_{\rm UE} = 16$}

\addplot [color=mycolor6, dashed, line width=1.2pt, mark=diamond, mark options={solid, mycolor6}]
  table[row sep=crcr]{%
5	1.55358125\\
10	1.55358125\\
20	1.55358125\\
40	1.55358125\\
80	1.55358125\\
160	1.55358125\\
};
\addlegendentry{$ M_{\rm gNB} = 64, M_{\rm UE} = 1$  $(\mbox{omni})$}

\end{axis}
\end{tikzpicture}%
    \caption{$T_{IA}$ as a function of $T_{SS}$ for the downlink configuration with analog gNB and hybrid UE. $N_{\rm SS} = 64$}
    \label{fig:tss64}
\end{figure}
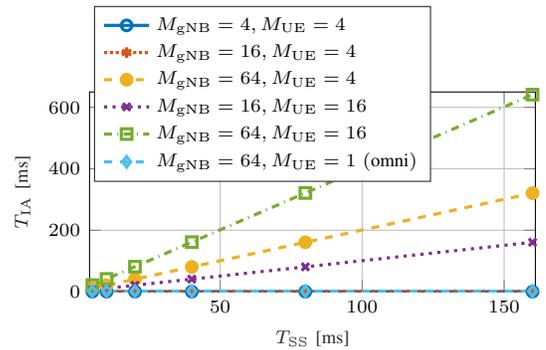


Furthermore,   Fig.~\ref{fig:tss64} shows the dependency of $T_{\rm IA}$ on $T_{\rm SS}$.  
It can be seen that the highest periodicities are not suited for a mmWave deployment, and that in general it is better to increase the number of \gls{ss} blocks per burst in order to try to complete the sweep in a single burst.



\textit{Impact of Beam Reporting.} For \gls{ia}, in addition to  the time required for  directional sweeping, there is also a delay related to the allocation of the resources in which it is possible to perform \gls{ia}, which differs according to the architecture being used.
As introduced in Sec. \ref{sec:meas_frameworks}, the 3GPP advocates the implicit reporting of the chosen direction, e.g., the strongest SS block index, through contention-based random access messages, agreeing that  the network should allocate multiple \gls{rach} transmissions and preambles to the UE for conveying the optimal SS block index to the gNB \cite{ericsson2017rach}.
When considering an SA configuration, beam reporting might require an additional sweep at the gNB side while, if an MC architecture is preferred, the beam decision is forwarded through the LTE interface and requires just a single \gls{rach} opportunity, which makes the beam reporting reactiveness equal to the latency of a legacy LTE connection.
Assuming no retransmissions are needed, the uplink latency in legacy LTE, including scheduling
delay, ranges from 0.8 ms to 10.5 ms, according to the latency reduction techniques being implemented \cite{latencyreduction2017}.

\begin{table}[b]
\centering
\footnotesize
\renewcommand{\arraystretch}{0.9}
\setlength\belowcaptionskip{-0.08cm}
\begin{tabular}{@{}lcccc@{}}
\toprule
& \multicolumn{4}{c}{$T_{\rm BR,SA}$ [ms]}                                       \\ 
                             & \multicolumn{2}{c}{$N_{\rm SS}=8$}      & \multicolumn{2}{c}{$N_{\rm SS}=64$}      \\
$M_{gNB}$                & Analog                 & Digital               &  Analog                 &  Digital                \\ \midrule
4                             & 0.0625               & 0.0625             & 0.0625               & 0.0625              \\
16                            & 0.5               & 0.0625             & 0.5              & 0.0625              \\
64                            & 40.56               & 0.0625             & 1.562               & 0.0625              \\ \midrule
\multicolumn{5}{c}{$T_{\rm BR,MC}=\{ 10, 4, 0.8\}$ [ms], according to \cite{latencyreduction2017}.}  \\ \bottomrule
\end{tabular}
\caption{Reactiveness performance for beam reporting operations considering an SA or an MC architecture. Analog or digital beamforming is implemented at the gNB side, while the UE configures its optimal beamformed direction. $T_{\rm SS}=20$ ms, $\Delta_{f}=120$ KHz.}
\label{tab:react_BR}
\end{table}

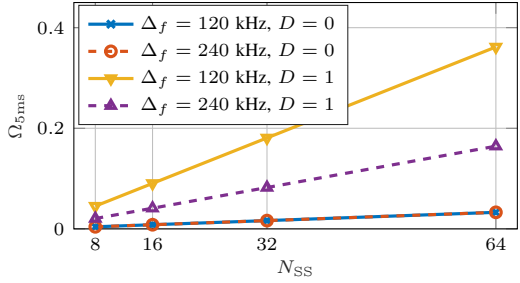
\begin{figure}[t!]
  \centering
  \setlength\belowcaptionskip{-0.6cm}
    \setlength\fwidth{0.7\columnwidth}
    \setlength\fheight{0.34\columnwidth}
%
%
\definecolor{mycolor1}{rgb}{0.00000,0.44700,0.74100}%
\definecolor{mycolor2}{rgb}{0.85000,0.32500,0.09800}%
\definecolor{mycolor3}{rgb}{0.92900,0.69400,0.12500}%
\definecolor{mycolor4}{rgb}{0.49400,0.18400,0.55600}%
\begin{tikzpicture}
\pgfplotsset{every tick label/.append style={font=\scriptsize}}

\begin{axis}[%
width=0.951\fwidth,
height=\fheight,
at={(0\fwidth,0\fheight)},
scale only axis,
xtick=data,
xmin=5,
xmax=67,
xlabel style={font=\scriptsize\color{white!15!black}},
xlabel={$N_{\rm SS}$},
ymin=0,
ymax=0.45,
ylabel style={font=\scriptsize\color{white!15!black}},
ylabel={$\Omega_{\rm 5ms}$},
ylabel shift=-2pt,
xlabel shift=-3pt,
axis background/.style={fill=white},
xmajorgrids,
ymajorgrids,
legend style={font=\scriptsize, at={(0.01, 0.99)}, anchor=north west, legend cell align=left, align=left, draw=white!15!black}
]
\addplot [color=mycolor1, line width=1.2, mark=x, mark options={solid, mycolor1}]
  table[row sep=crcr]{%
8	0.00410976\\
16	0.00821952\\
32	0.01643904\\
64	0.03287808\\
};
\addlegendentry{$\Delta_f = 120$~kHz, $D = 0$}

\addplot [color=mycolor2, line width=1.2, dashed, mark=o, mark options={solid, mycolor2}]
  table[row sep=crcr]{%
8	0.00410976\\
16	0.00821952\\
32	0.01643904\\
64	0.03287808\\
};
\addlegendentry{$\Delta_f = 240$~kHz, $D = 0$}

\addplot [color=mycolor3, line width=1.2, mark=triangle, mark options={solid, rotate=180, mycolor3}]
  table[row sep=crcr]{%
8	0.04520736\\
16	0.09041472\\
32	0.18082944\\
64	0.36165888\\
};
\addlegendentry{$\Delta_f = 120$~kHz, $D = 1$}

\addplot [color=mycolor4, line width=1.2, dashed, mark=triangle, mark options={solid, mycolor4}]
  table[row sep=crcr]{%
8	0.0205488\\
16	0.0410976\\
32	0.0821952\\
64	0.1643904\\
};
\addlegendentry{$\Delta_f = 240$~kHz, $D = 1$}

\end{axis}
\end{tikzpicture}%
  \caption{Overhead $\Omega_{\rm 5ms}$ for \gls{ia} as a function of $N_{\rm SS}$, introduced by the transmission of the \gls{ss} blocks. Notice that the number of repetitions for the different subcarrier spacings $\Delta_f$ is chosen to send as many repetitions of the \gls{ss} blocks as possible.}
  \label{fig:omega_ia}
\end{figure}

In Table \ref{tab:react_BR}, we  analyze the impact of the number of SS blocks (and, consequently, of RACH opportunities) in a burst, with a fixed burst periodicity $T_{\rm SS} = 20$ ms and for a subcarrier spacing of $\Delta_{f}=120$ KHz.
The results are independent of the antenna configuration at the UE side, since the mobile terminal steers its beam through the previously determined optimal direction and does not require a beam sweeping operation to be performed.
It appears clear that the SA scheme presents very good reactiveness for most of the investigated configurations and, most importantly, outperforms the MC solution even when the LTE latency is reduced to $0.8$ ms.
The reason is that, if the network is able to allocate the needed RACH resources within a single SS burst, then it is possible to limit the impact of beam reporting operations on the overall \gls{ia} reactiveness, which is dominated by the beam sweeping phase instead.
In particular, when considering small antenna factors and when digital beamforming is employed, beam reporting can be successfully completed through a single RACH allocation, thus guaranteeing very small delays.

\subsection{Overhead Results}
\label{sec:overhead}
In this section, we characterize the overhead for \gls{ia} in terms of the ratio between the time and frequency resources that are allocated to \gls{ss} bursts and the maximum duration of the \gls{ss} burst (i.e., 5 ms), or the entire $T_{\rm SS}$ interval. 

The total number of time and frequency resources $R_{\rm SS}$ scheduled for the transmission of $N_{\rm SS}$ \gls{ss} blocks, each spanning 4 \gls{ofdm} symbols and 240 (or a multiple of 240) subcarriers, is given by $R_{\rm SS} = N_{\rm SS} \; 4T_{\rm symb} \; 240N_{rep}\Delta_f$,
where $T_{\rm symb}$ is expressed in ms.
The overhead for the 5 ms time interval in which the \gls{ss} burst transmission happens, and with total bandwidth $B$, is then given by
\begin{equation}
	\Omega_{\rm 5ms} = \frac{N_{\rm SS} \; 4T_{\rm symb} \; 240N_{rep}\Delta_f}{5 B},
\end{equation}
while the overhead considering the total burst periodicity $T_{\rm SS}$~is
\begin{equation}\label{eq:ovTss}
	\Omega_{\rm T_{\rm SS}} = \frac{N_{\rm SS} \; 4T_{\rm symb} \; 240N_{rep}\Delta_f}{T_{\rm SS} B}.
\end{equation}

Fig.~\ref{fig:omega_ia} reports the overhead related  to the maximum duration of the \gls{ss} burst (i.e., 5 ms) for different subcarrier spacings and repetition strategies. It can be seen that if no repetitions are used (i.e., $D=0$) then the overheads for the configurations with $\Delta_f = 120$~kHz and $\Delta_f = 240$~kHz are equivalent: the \gls{ofdm} symbols used for the \gls{ss} blocks have  half the duration with the larger subcarrier spacing, but they occupy twice the bandwidth, given that the same number of subcarriers are used. Instead, when a repetition strategy is used (i.e., $D=1$), the overhead is higher. As mentioned in Sec.~\ref{sec:system_model}, we consider 5 repetitions for $\Delta_f = 240$ kHz and 11 for $\Delta_f = 120$ kHz. Therefore, the actual amount of bandwidth that is used is comparable, but since the \gls{ofdm} symbols with $\Delta_f= 120$~kHz last twice as long as  those with the larger subcarrier spacing, the overhead in terms of resources used for the \gls{ss} burst is higher with $\Delta_f= 120$~kHz. 
%
%
%
\begin{table}[t!]
\centering
\footnotesize
  \renewcommand{\arraystretch}{1}
\begin{tabular}{lcccc}
\toprule
& \multicolumn{4}{c}{$\Omega_{\rm BR,SA}$ $\cdot10^{-3}$}                                       \\ 
                             & \multicolumn{2}{c}{$	\Delta_{f,\rm RACH}=60$ kHz}      & \multicolumn{2}{c}{$	\Delta_{f,\rm RACH}=120$ kHz}      \\
$M_{gNB}$                             & Analog                  & Digital                & Analog                  & Digital                 \\ \midrule
\multicolumn{1}{c}{4}                             & 0.0894               & 0.0894              &  0.0894               & 0.0894           \\
\multicolumn{1}{c}{16}                            &  0.7149              & 0.0894              & 0.7149              & 0.0894            \\
\multicolumn{1}{c}{64}                            & 2.2341               & 0.0894              & 2.2341               & 0.0894           \\ \bottomrule
\end{tabular}
\caption{Overhead for beam reporting operations considering an SA architecture. Analog or digital beamforming is implemented at the gNB side, for different antenna array structures.}
\label{tab:ov_BR}
\end{table}

\emph{Impact of beam reporting}\footnote{According to the 3GPP agreements \cite{38211}, a bandwidth of about $10$ MHz or $20$ MHz is reserved for the RACH resources, respectively for $\Delta_{f,\rm RACH}=60$ kHz or $120$ kHz.}.
 For the SA case, as reported in Table \ref{tab:ov_BR}, the completion of the beam reporting procedure for \gls{ia} may require additional overhead, due to the need for the system to allocate possibly multiple RACH resources for the reporting operations.
 Conversely, for the MC case, the beam decision is forwarded through the LTE overlay and requires
a single RACH opportunity, with a total overhead of $0.0894\cdot10^{-3}$. 
Nevertheless, from Table \ref{tab:ov_BR}, we
notice that the SA additional reporting overhead  is quite limited due to the relatively small number of directions that need to be investigated at this stage, especially when designing digital beamforming solutions.

\subsection{Final Considerations}
\label{sec:final_consid}
Overall, it is possible to identify some guidelines for the configuration of the \gls{ia} framework and the deployment of an NR network at mmWave frequencies. First, a choice of  $N_{\rm SS}$, the \gls{rach} resources, the beamforming and the antenna array architectures that allows the completion of the beam sweeping and reporting procedures in a single burst is preferable, so that it is possible to increase $T_{\rm SS}$ (e.g., to 20 or 40 ms), thereby reducing the overhead of the \gls{ss}~blocks. 

Second, the adoption of a frequency diversity scheme increases the detection accuracy at the expense of an increased overhead.
Nevertheless, the accuracy gain decreases
when the antenna array dimension is increased: in those
circumstances, it may not be desirable to adopt a frequency diversity scheme which would lead to limited performance improvements.

Third, with low network density, larger antenna arrays can reach farther users and provide a wider coverage but, as $\lambda_b$ increases, it is possible to use a configuration with wide beams for \gls{ss} bursts (so that it is more likely to complete a sweep in a single burst) during \gls{ia} and narrow ones for data transmission, to refine the pointing directions and achieve higher gains.

Fourth, when considering stable and dense scenarios which are marginally affected by the variability of the mmWave channel, a standalone architecture is preferable for the design of fast \gls{ia} procedures, since it enables rapid beam reporting operations, at the expense of a slightly increased overhead.
Anyway,  we still claim that an \gls{mc} configuration may be preferable for several other reasons, including:
\begin{itemize}
\item[(i)] \emph{Reduced overhead:} the impact of the reporting operations on the communication performance at mmWave frequencies is almost negligible.
\item[(ii)] \emph{Successful beam reporting:} an MC scheme eliminates the need for the UE to send measurement feedback through mmWave connections (which are much more volatile than their LTE-based counterpart)
and thereby removes a possible point of failure in the control signaling path.
\item[(iii)] \emph{Centralized beam decision:} unlike in traditional attachment policies based on pathloss measurements, by leveraging on the
presence of an eNB operating at sub-6 GHz frequencies, an MC-based initial association can be possibly performed by taking
into account the instantaneous load conditions of the surrounding cells, thereby promoting
fairness in the whole cellular network \cite{giordani2016efficient}.
\end{itemize}

Finally, a downlink configuration is in line with the 3GPP design for NR and reduces the energy consumption at the \gls{ue} side (since it has just to receive the synchronization or reference signals), but  is less reactive because the \glspl{gnb} have a larger number of directions to sweep with downlink \gls{ss} blocks. 

\vspace{-.1cm}
\section{Conclusions}
\label{sec:concl}
\vspace{-.1cm}
The extreme propagation environment at mmWave frequencies requires the adoption of directional transmissions and beamforming techniques, which increase the achievable data rate but also the latency and overhead required to perform \gls{ia}. 
In this paper we evaluated, with an extensive analysis and simulation
campaign, the impact of several parameters (specified by the 3GPP for NR) on 
the performance of multiple \gls{ia} schemes for NR networks operating at mmWaves.
We showed that there exist tradeoffs among better detection accuracy, improved reactiveness and reduced overhead.
We therefore provided  guidelines for determining the optimal \gls{ia} strategy in different network deployments, according to the needs of the network operator and the specific environment in which the nodes are~deployed. 

As part of our future work, we will extend the analysis to the tracking of the beam quality for users in connected state (i.e., users that have successfully completed \gls{ia}), and investigate the resilience performance of the  beam management frameworks when considering radio link failure, outage~events and the impact of LTE retransmissions. 

\vspace{-.2cm}
\bibliographystyle{IEEEtran}
\bibliography{bibl_medhoc.bib}

\end{document}